 \documentstyle[11pt]{article}
 \newcommand{\be}{\begin{equation}}
 \newcommand{\ee}{\end{equation}}
 \newcommand{\bea}{\begin{eqnarray}}
 \newcommand{\eea}{\end{eqnarray}}

\begin{document}
 \begin{center}
 {\LARGE\bf On the detectability of post-Newtonian effects in
 gravitational-wave emission of a coalescing binary}\footnote{This work was
 supported by KBN Grant No. 2 P302 076 04}
 \end{center}
 \vspace{6mm}

 ANDRZEJ KR\'OLAK$^a$ \hspace{3mm} KOSTAS D. KOKKOTAS$^b$  \hspace{3mm}
 GERHARD SCH\"AFER$^c$

 \vspace{2mm}

 \begin{center}
 $^a$Institute of Mathematics \\
 Polish Academy of Sciences \\
 \'Sniadeckich 8, 00-950 Warsaw, Poland

 \vspace{1mm}

 $^b$Department of Physics, Section Astrophysics, Astronomy and Mechanics, \\
 Aristotle University of Thessaloniki \\
 540 06 Thessaloniki, Macedonia, Greece

 \vspace{1mm}

 $^c$Max-Planck-Research-Group Gravitational Theory \\
 at the Friedrich-Schiller-University \\
 07743 Jena, Germany
 \end{center}

 \vspace{8mm}

 \centerline{\bf INTRODUCTION}

 \vspace{4mm}

 It is currently believed that the gravitational waves
 that come from the final
 stages of the evolution of compact binaries just before their coalescence
 are very likely signals to be detected by long-arm laser interferometers
 \cite{T}.
 The two projects to build such detectors -  LIGO
 and VIRGO - are already approved and they are rapidly progressing, and
the third one - GEO600 - is likely to be funded soon.
 A standard optimal method to detect the signal from a coalescing binary
 in a noisy data set and to estimate its parameters is to correlate
 the data with the filter matched to the signal and vary the parameters of the
 filter until the correlation is maximal.
 It has recently been realized$^{\cite{Cal}}$ that the correlation is very
 sensitive even to very small variations of the phase of the filter because
 of the large number of cycles in the signal.
 Consequently, the addition of small corrections to the phase of the
  signal due
 to the post-Newtonian effects decreases the correlation considerably.
 Thus the post-Newtonian effects in the coalescing binary waveform can
 be detected and estimated to a much higher accuracy than it was thought
 before$^{\cite{K}}$.
 We present an analysis of the estimation of parameters of the
 post-Newtonian signal. We also examine the
 detectability of the post-Newtonian signal and estimation of its parameters
 using the Newtonian waveform as a filter. This filter can be used
 as the simplest search template.

 \vspace{4mm}

 \centerline{\bf GRAVITATIONAL-WAVE SIGNAL FROM A BINARY}

 \vspace{4mm}

 Let us first give the formula for the gravitational waveform of a binary
 with the currently known post-Newtonian corrections.
 In this paper we work within the so called ``restricted" post-Newtonian
 approximation, i.e., we only include the post-Newtonian
 corrections to the phase of the signal keeping the amplitude in its
 Newtonian form; this is because the effect of the phase on
 the correlation is dominant. The inclusion of post-Newtonian effects
 in amplitudes will not qualitatively change our results.
 Due to radiation reaction the orbit of the binary is rapidly circularized,
 nevertheless, we include the first order correction due to eccentricity
 for completeness\footnote{This correction was derived by N. Wex}.
 The tidal effects are known to be very small and we neglect them.
 We also include the contributions due to dipole radiation
 predicted by the Jordan-Fierz-Brans-Dicke (JFBD) theory (assuming that they
 are small) to investigate the possibility of testing GR against alternative
 theories of gravity.

 The analysis of the signal is best performed in the Fourier domain.
 The expression for the Fourier transform of our signal in the stationary phase
 approximation is given by
 \be
 \tilde{s} = \frac{1}{(30)^{1/2}}\frac{1}{\pi^{2/3}}
             \frac{\mu^{1/2}m^{1/3}}{R}\tilde{h}
 \ee
 where
 \bea
 & & \tilde{h} = f^{-7/6} \exp i[2\pi f t_c - \phi_c - \pi/4  - \\ \nonumber
 & &\frac{7065}{187136}\frac{k_e}{(\pi f)^{34/9}}
 + \frac{3}{128}\frac{k}{(\pi f)^{5/3}}
 + \frac{5}{96}\frac{k_1}{(\pi f)^{1}}
 - \frac{3}{32}\frac{k_{3/2}}{(\pi f)^{2/3}}
 + \frac{3}{128}\frac{k_2}{(\pi f)^{1/3}} \\ \nonumber
 & & - \frac{5}{14336}\frac{k_D}{(\pi f)^{7/3}}],
 \label{Fou}
 \end{eqnarray}
 (for $f > 0$, and by the complex conjugate of the above expression,
 for $f < 0$) where
 \bea
 k &=& \frac{1}{\mu m^{2/3}}(1 -
 \frac{F({\cal C}_1, {\cal C}_2, m_1, m_2)}{2 + \omega}),\
 k_e = \frac{1}{\mu m^{2/3}}e_o^2(\pi f_o)^{19/9} ,\\ \nonumber
 k_1 &=& \frac{1}{\mu}(\frac{743}{336} + \frac{11}{4}\frac{\mu}{m}),\
 k_{3/2} = \frac{m^{1/3}}{\mu}(4\pi - s_o), \\ \nonumber
 k_2 &=& \frac{m^{4/3}}{\mu}(3.0 + c_1\frac{\mu}{m} + c_2(\frac{\mu}{m})^2 +
 s_s\frac{m}{\mu}) ,\\ \nonumber
 k_D &=& \frac{1}{\mu m^{4/3}}\frac{({\cal C}_1 - {\cal C}_2)^2}{2 + \omega}
 \eea
 where $e_o$ is the eccentricity of the binary at gravitational frequency
 $f_o$ ($k_e$ is constant to order $e_o^2$), $s_o$ and $s_s$ are
spin-orbit and spin-spin parameters,  respectively$^{\cite{WWW}}$.
 The constants $c_1$ and $c_2$ in the 2nd post-Newtonian contribution are
 only recently being calculated \cite{BW} and calculation of further
 corrections is in progress.
 ${\cal C}_i, i$=(1,2), is the sensitivity of the body $i$ to changes
 of the scalar field, $F({\cal C}_1, {\cal C}_2, m_1, m_2)$
 is a complicated function of sensitivities and masses of the bodies
 and $\omega$ is the parameter of the JFBD theory. Current observational
 tests restrict $\omega > 600$ and observations of the Hulse-Taylor
 binary pulsar restrict $\omega > 200$.

 \vspace{4mm}

 \centerline{\bf ESTIMATION OF THE POST-NEWTONIAN PARAMETERS}

 \vspace{4mm}

 A basic method proposed to detect the gravitational-wave signal
 of a coalescing binary and to estimate its parameters is matched-filtering and
 maximum likelihood (ML) estimation. The performance of the method
 is determined by signal-to-noise ratio d and the Fisher information matrix
 $\Gamma$. Signal-to-noise ratio determines the probability of detection
 of the signal and the diagonal elements of the
 inverse of the Fisher matrix are approximately the variances of the
 maximum likelihood estimators of the parameters for large d.

 We have considered three representative binary systems with parameters
 summarized in Table I.

 {\bf Table I.} Parameters of Neutron Star (NS) and Black-Hole (BH)
 Binary Systems

 \begin{tabular}{|c|c|c|c|c|c|c|c|} \hline
 Binary & $m_1 M_{\odot}$ & $m_2 M_{\odot}$ & ${\cal M} M_{\odot}$ & s
 & $k M_{\odot}^{-5/3}$ & $k_1 M_{\odot}^{-1}$ & $k_{3/2}
M_{\odot}^{-2/3}$ \\ \hline
 NS-NS  & 1.4 & 1.4 & 1.2 & 2.4$\times 10^{-2}$ & 0.72 & 4.1 & 25 \\
 NS-BH  & 1.4 & 10 & 3.0  & 4.0 & 0.16 & 2.0 & 16 \\
 BH-BH  & 10 & 10 & 8.7 & 3.9 & 2.7$\times10^{-2}$ & 0.58 & 4.7 \\ \hline
 \end{tabular}

 \bigskip

 Assuming that the noise is Gaussian, neglecting the effects of eccentricity,
 2nd post-Newtonian effects, dipole radiation, and assumining that
binaries are
 located at a distance of 200Mpc we have calculated the signal-to-noise ratios
 and approximate values of the variances of the ML estimators.
 The results are summarized in Table II (cf. \cite{CF} Table II).

 {\bf Table II.} Signal-to-Noise Ratio and RMS Errors for Parameters
 of the 3 Binaries

 \begin{tabular}{|c|c|c|c|c|c|c|c|c|} \hline
 Binary & S/N & $\sqrt{n}$ & $\Delta t_a$ms & $\Delta {\cal M}/{\cal M}$
 & $\Delta \mu/\mu$ & $\Delta m/m$ & $\Delta s/s$ \\ \hline
 NS-NS  & 15 & 32 & 0.75 & 0.023\% & 6.4\% & 9.6\% & $33 \times 10^2$\% \\
 NS-BH  & 32 & 15 & 0.65 & 0.061\% & 4.8\% & 7.0\% & 8.2\% \\
 BH-BH  & 77 & 6 & 0.46 & 0.19\% & 16\% & 23\% & 38\% \\ \hline
 \end{tabular}

 \bigskip

 The number $\sqrt{n}$ gives the improvement of the signal-to-noise ratio due
 to matched filtering.

 We have also examined the effects of eccentricity and dipole radiation
 We have added their contributions to the phase of the 3/2post-Newtonian signal
 separately and we have calculated the Fisher matrix and its inverse.
 For a NS-NS binary the relative rms error in eccentricity is given by
 $\frac{\Delta e}{e} = \frac{0.63 \times 10^{-6}}{a_e}$.
 For the Hulse - Taylor binary pulsar we have $a_e = e_o^2f_o^{19/9} =
 1.8 \times 10^{-13}$  .
 Thus the effects of eccentricity would be practically undetectable
 for such a binary.
 For the NS-BH binary the relative rms error in the dipole radiation
 coefficient $k_D$ in JFBD theory is given by
 $\frac{\Delta k_D}{k_D} = 0.75\left(\frac{\omega}{100}\right)$.
 Thus, the effects of the dipole radiation could be determined to about
 the same accuracy as from timing of the binary pulsar in our Galaxy.

 It is well known that the rms errors of the estimators of the parameters
 increase with the number of parameters. We have investigated this effect
 with the increasing number of post-Newtonian parameters.
 We have considered a reference binary  of ${\cal M} = 1M_{\odot}$ located
 at the distance of 100Mpc.

 \bigskip

 {\bf Table III} RMS Errors for Parameters at Various post-Newtonian Orders

 \hspace*{-15mm}
 \begin{tabular}{|c|c|c|c|c|c|c|} \hline
 $\Delta t_c$ msec & $\Delta\phi_c$ & $\Delta k M_{\odot}^{-5/3}$
& $\Delta k_1 M_{\odot}^{-1}$
 & $\Delta k_{3/2} M_{\odot}^{-2/3}$ & $\Delta k_2 M_{\odot}^{-1/3}$ &
 $\Delta k_e M_{\odot}^{-19/9}$ \\ \hline
 0.17 & 0.10 & 8.3 $\times 10^{-6}$ & - & - & - & - \\
 0.27 & 0.33 & 4.0 $\times 10^{-5}$ & 5.8$\times 10^{-3}$ & - & - & - \\
 0.54 & 1.9 & 1.7 $\times10^{-4}$ &  0.70 $\times10^{-1}$ & 0.52 & - & - \\
 1.6 & 24 & 6.6 $\times10^{-4}$ &  0.50 & 7.2 & 28 & - \\
 2.3 & 45 & 2.3 $\times10^{-3}$ &  1.3 & 17 & 59 & 1.2$\times10^{-6}$ \\ \hline
 \end{tabular}

 \bigskip

 We were able to include the 2nd post-Newtonian correction because the Fourier
 transform of the signal is linear in the mass parameters $k_i$ and
 consequently, the Fisher matrix for these parameters does not depend
 on their numerical values.
 In Table IV we show the degradation of accuracy of estimation
 of the chirp mass, the reduced mass, and the total mass with
 the increasing number
 of parameters in the template for the NS-NS binary at a distance of 200Mpc.

 \bigskip

 {\bf Table IV.}  RMS Errors for Masses at Various post-Newtonian Orders

 \begin{tabular}{|c|c|c|c|} \hline
 pN order & $\Delta {\cal M}/{\cal M}$ & $\Delta \mu/\mu$ & $\Delta m/m$ \\
\hline
 1 pN & 0.0054\% & 0.55\% & 0.81\% \\
 3/2 pN & 0.023\% & 6.4\% & 9.6\% \\
 2 pN & 0.080\% & 42\% & 63\% \\ \hline
 \end{tabular}

 \bigskip

 The rms errors in $\mu$ and $m$ do not depend on the value of 2nd
post-Newtonian
 mass paramter $k_2$.

 \vspace{4mm}

 \centerline{\bf THE NEWTONIAN FILTER}

 \vspace{4mm}

 On the one hand the correlation of the signal with the template is very
sensitive
 to small corrections in the phase of the signal on the other hand
 the accuracy of estimation of the parameters is significantly degraded
 with increasing number of corrections even though a correction may be small.
 Moreover we cannot entirely exclude unpredicted small effects
 in the gravitational-wave emmission
(e.g. corrections to general theory of gravity)
 that we present cannot model. Thus, there is a need for simple filters
 or {\em search templates} that will unable to scan the data effectively
 and isolate stretches of data where the signal is most likely
 to be$^{\cite{Cal}}$. The simplest such filter is just a Newtonian waveform
 $h_N$ the Fourier transform of which, in stationary phase approximation, is
given by   $\tilde{h}_N = f^{-7/6} \exp i[2\pi f t_c - \phi_c - \pi/4 +
 k\frac{3}{128}(\pi f)^{-5/3}$]. Using such a suboptimal filter decreases
 the correlation and, consequently, decreases the signal-to-noise ratio. Also
 the estimate of the k parameter is shifted by a definite amount depending
 on the noise of the detector and the parameters of the binary. In Table V
 we have given the drop in signal-to-noise given by $FF = (h|h_N)/(h|h)$
 where $(h_1|h_2)$ is the correlation  of functions $h_1$ and $h_2$
 (see Ref.\cite{CF} Eq.(2.3) for definition) and the parameter $l = \sqrt{FF}$.
 In the case of the Newtonian filter probability of detection is determined by
 two parameters: optimal signal-to-noise ratio $d = \sqrt{(h|h)}$ and
 $d_o = \sqrt{(h|h_o)} = l \times d$. We have also calculated the shift
 in the estimate of the k parameter and the accuracy of the determination of k.
 We have included the 1st and the 3/2 post-Newtonian corrections.

 {\bf Table V.} Performance of the Newtonian Filter

 \begin{tabular}{|c|c|c|c|c|} \hline
 Binary &  $l$ & FF &
 Shift  $\delta k M_{\odot}^{-5/3}$ &
 Accuracy $\Delta k M_{\odot}^{-5/3}$ \\ \hline
 NS-NS & 0.90 & 0.81 & 0.01560 & $0.13 \times 10^{-3}$ \\
 NS-BH & 0.87 & 0.76 & 0.004898 & $0.026 \times 10^{-3}$\\
 BH-BH & 0.98 & 0.96 & 0.001209 & $0.011 \times 10^{-3}$\\ \hline
 \end{tabular}

 \bigskip

 When the amplitude and phase modulations are taken into account it was
 shown$^{\cite{Ap}}$ that in the worst case FF = 0.39 (l = 0.63).

 Using the Newtonian filter we would not like to loose any signals.
 We can achieve this by suitably lowering the detection
 threshold when filtering the data
 with the Newtonian filter.
By this procedure we would isolate stretches of data
 with all the signals that would be detected with optimal filter and also
 an increased number of false alarms. The next step would be to analyse
 the reduced set of data with accurate templates and the initial
 threshold to make the final detection and estimate the
 parameters of the signal.

 In Table VI we have given examples of the performance of the above
procedure. We assume the detection threshold T = 5 and
 we assume that we have one signal for the optimal signal-to-noise ratio d.
 N is the expected number of detected signals with the optimal filter,
 $N_F$ is number of false alarms,
 $N_N$ is the number of detected signals with the Newtonian filter,
 $T_N$ is the lowered threshold, $N_L$ is number of signals with the lowered
 threshold  and $N_{FL}$ is number of false alarms with the lowered threshold.

 {\bf Table VI}

 \begin{tabular}{|c|c|c|c|c|c|c|c|} \hline
 d & FF & N & $N_F$ & $N_N$ & $T_N$ & $N_L$ & $N_{FL}$ \\ \hline
 15 & .81 & 27 & 0.055 & 20 & 4.5 & 28 & 0.16 \\
 15 & .36 & 27 & 0.055 & 5.6 & 3.225 & 28 & 2.1 \\
 30 & .81 & 225 & 1.1 & 165 & 4.5 & 230 & 2.2 \\
 30 & .25 & 225 & 1.1 & 31 & 2.875 & 229 & 32 \\ \hline
 \end{tabular}

 \vspace{3mm}

 A different search template consisting of the post-Newtonian waveform with
 spin effects stripped off has been analysed in Ref.\cite{Ap}.

 \vspace{4mm}

 \centerline{ACKNOWLEDGEMENTS}

 \vspace{4mm}

 A.K. thanks the Max-Planck-Gesellschaft
 for support and the Arbeitsgruppe Gravitationstheorie an der Friedrich-
 Schiller-Universit\"at in Jena for hospitality during the time this
 work was done. We would like to thank T. Apostolatos and K.S. Thorne
 for helpful discussions.

 \end{document}